\documentclass[12pt]{article}
\usepackage{graphicx}

\newcommand\pubdate{\today}

\def\Title#1{\begin{center} {\Large #1 } \end{center}}
\def\Author#1{\begin{center}{ \sc #1} \end{center}}
\def\Address#1{\begin{center}{ \it #1} \end{center}}

\newcommand\pubblock{\rightline{\begin{tabular}{l}  \\ 
         \pubdate  \end{tabular}}}
\newenvironment{Abstract}{\begin{quotation}  }{\end{quotation}}
\newenvironment{Presented}{\begin{quotation} \begin{center} 
             PRESENTED AT\end{center}\bigskip 
      \begin{center}\begin{large}}{\end{large}\end{center} \end{quotation}}

\begin{document}
\begin{titlepage}
 \pubblock
\vfill
\Title{Jet separated by a large rapidity gap at the Tevatron and the LHC }
\vfill
\Author{Christophe Royon}
\Address{Department of Physics and Astronomy, The University of Kansas, Lawrence, USA}
\vfill
\begin{Abstract}
We compare the recent measurements of gap between jets at the Tevatron and the LHC with the Balitski Fadin Kuraev Lipatov framework. While a good agreement is obtained with Tevatron data, some discrepancies especially for the rapidity separation between jets are found that can be explained by an excess of initial state radiation in PYTHIA.
\end{Abstract}
\vfill
\begin{Presented}
DIS2023: XXX International Workshop on Deep-Inelastic Scattering and
Related Subjects, \\
Michigan State University, USA, 27-31 March 2023 \\
     \includegraphics[width=9cm]{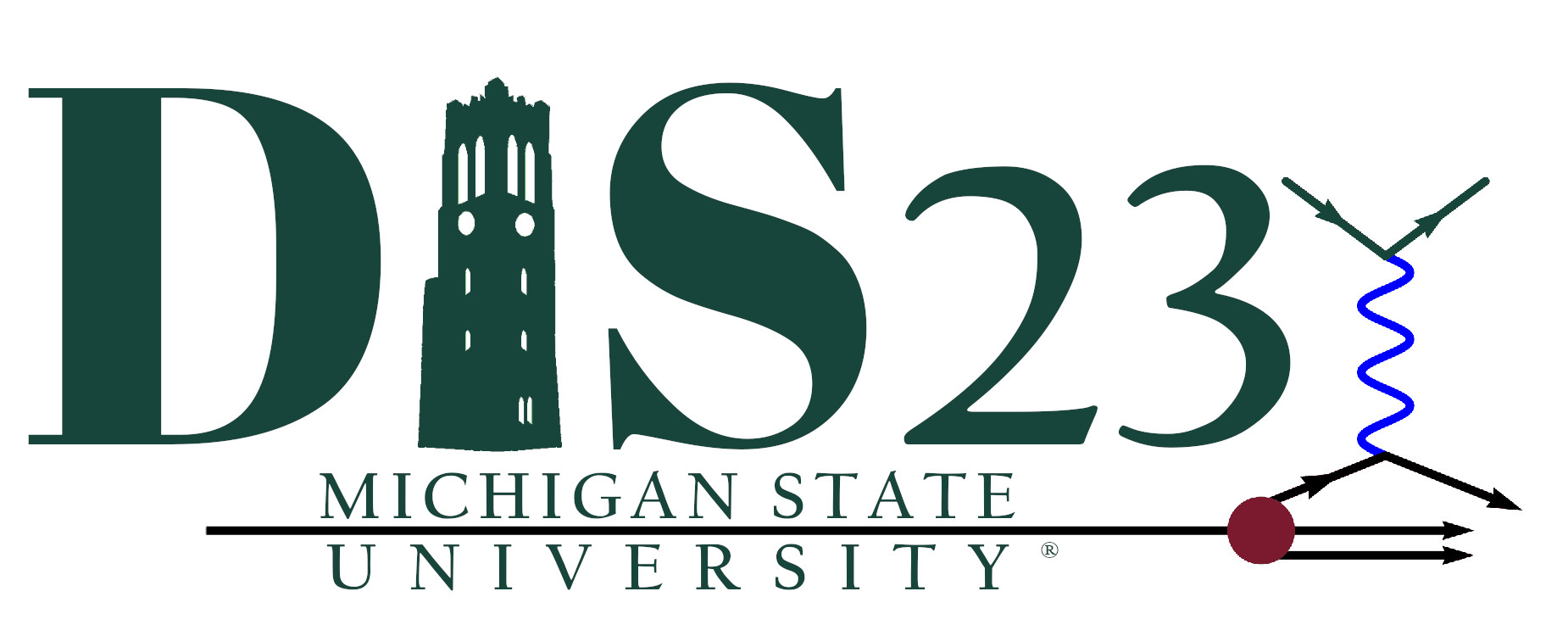}
\end{Presented}
\vfill
\end{titlepage}

\section{Introduction and BFKL formalism}

\begin{figure}
\centerline{%
\includegraphics[width=11.cm]{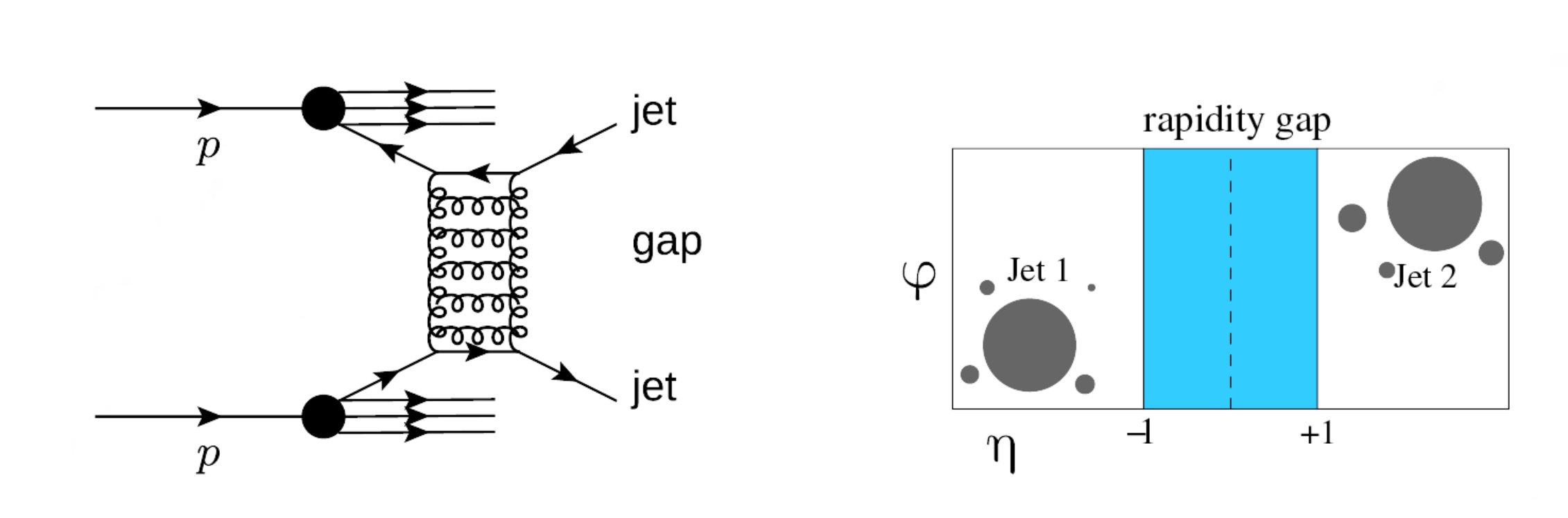}}
\caption{Schematic of jet gap jet events.}
\label{fig1}
\end{figure}

In this paper, we discuss the description of recent measurements of gap between jets, the so-called Mueller-Tang process~\cite{mt} at the Tevatron and the LHC using the Balitski Fadin Kuraev Lipatov (BFKL)~\cite{bfkl} formalism. The schematic of jet gap jet events is shown in Fig.~\ref{fig1}. Two jets separated by a difference in rapidity $\Delta \eta$ are measured in the detector while a region in rapidity between (-1) and (1) is devoid of any particle. Experimentally, it is possible to veto on the presence of energy inside the calorimeter or on tracks from charged particles. A Pomeron is exchanged between the two jets so that there is no color flow. The natural dynamics to describe this kind of events is the BFKL one while the Dokshitzer Gribov Lipatov Altarelli Parisi (DGLAP)~\cite{dglap} formalism leads to a negligible cross section when gaps are large enough, typically more than 1.5 units of rapidity.

The BFKL jet gap jet cross section reads
\begin{eqnarray}
\frac{d \sigma^{pp\to XJJY}}{dx_1 dx_2 dp_T^2} = 
{\cal S}\frac{f_{eff}(x_1,p_T^2)f_{eff}(x_2,p_T^2)}{16 \pi}
\left|A(\Delta\eta,p_T^2)\right|^2 
\end{eqnarray}
where $p_T$ is the jet transverse momentum (we assume that we have only two jets of same $p_T$ that are produced at parton level), $\Delta \eta$ the separation in rapidity between the two jets, $x_1$ and $x_2$ the energy fraction carried away by the jet, and $S$ the survival probability (0.1 at Tevatron, 0.03 at LHC). The amplitude $A$ reads
\begin{eqnarray}
A=\frac{16N_c\pi\alpha_s^2}{C_Fp_T^2}\sum_{p=-\infty}^\infty
\int \frac{d \gamma}{2 i \pi}
\frac{[p^2-(\gamma-1/2)^2]}{[(\gamma-1/2)^2-(p-1/2)^2]}  \nonumber 
 \frac{\exp\left\{\frac{\alpha_S N_C}{\pi} 
\chi_{eff} \Delta \eta\right\}}
{[(\gamma-1/2)^2-(p+1/2)^2]}
\end{eqnarray}
where the sum stands over all conformal spins, 
$\alpha_S$ is constant at LL and running using the renormalization group equations at NLL.
The BFKL effective kernel $\chi_{eff}$ is
determined numerically, solving the implicit equation
$\chi_{eff}=\chi_{NLL}(\gamma,\bar\alpha\ \chi_{eff})$~\cite{mtus}.
The S4 resummation scheme~\cite{salam} is used to remove spurious singularities 
in the BFKL NLL kernel.
This formalism was fully Implemented in the HERWIG~\cite{herwig}  and PYTHIA~\cite{pythia} Monte Carlos, which is needed to take
into account the jet size and the fact that the gap size is smaller than $\Delta \eta$ between the jets by definition, the gap being defined at the edge of the jets~\cite{mtus}.

\section{Comparison between the BFKL prediction and the measurement at the Tevatron and the LHC}
 
\begin{figure}
\centerline{%
\includegraphics[width=9.cm]{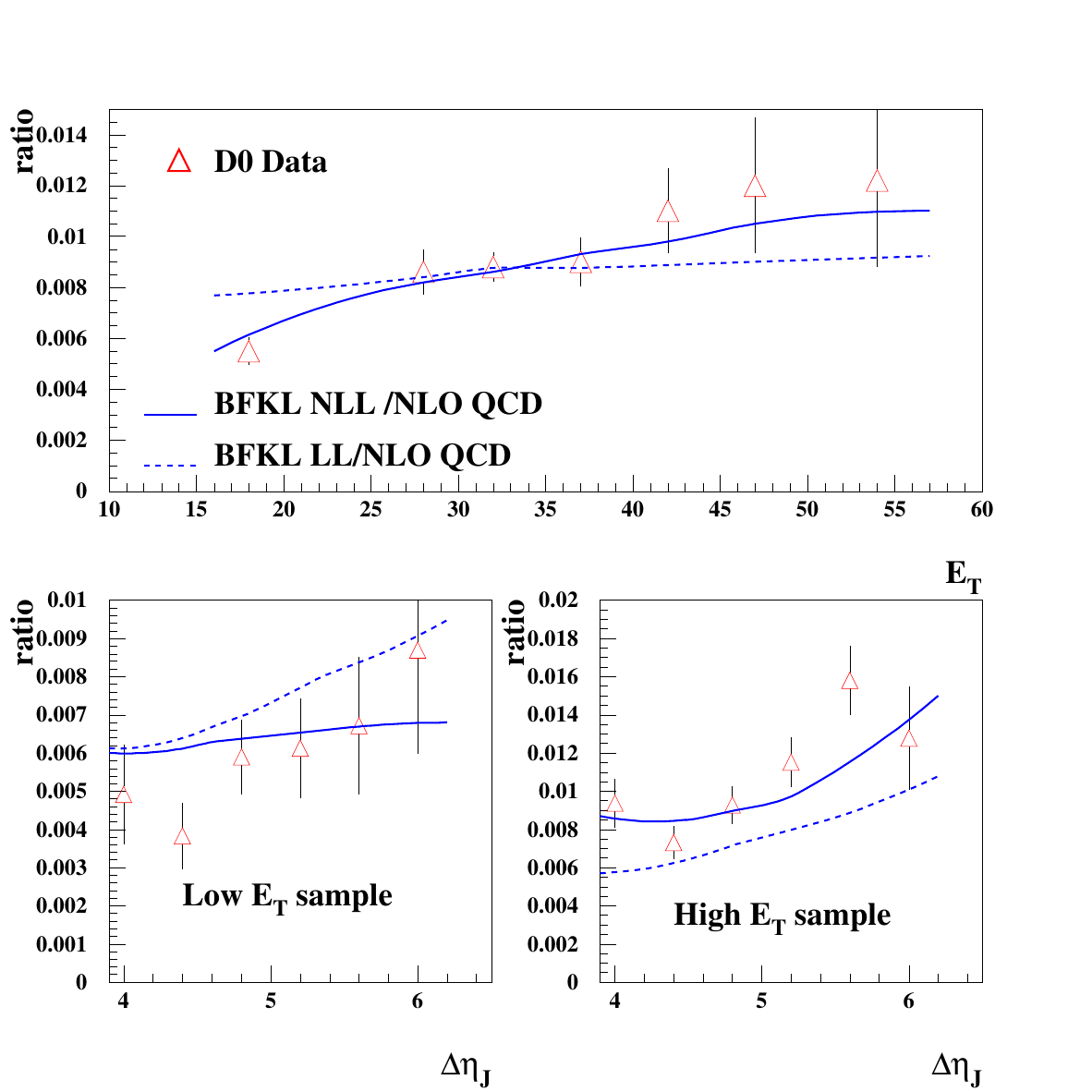}}
\caption{Measurement of jet gap jet events at the Tevatron by the D0 collaboration compared with the BFKL LL and NLL calculation.}
\label{fig2}
\end{figure}

\begin{figure}
\centerline{%
\includegraphics[width=16.cm]{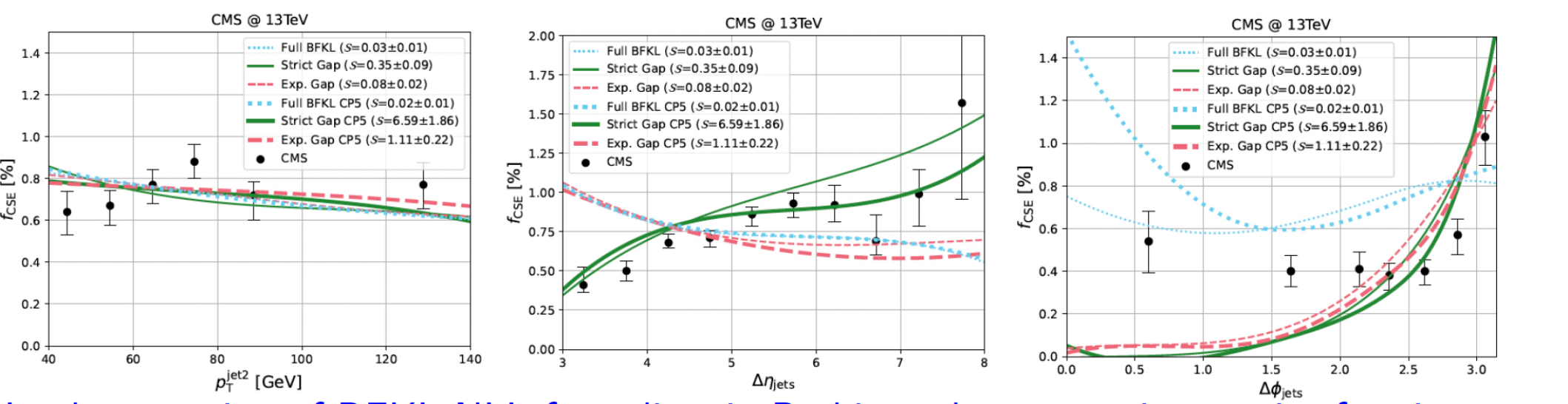}}
\caption{Measurement of jet gap jet events at the LHC by the CMS collaboration compared with the BFKL  NLL calculation with different gap definitions (BFKL, experimental and strict).}
\label{fig3}
\end{figure}

 The D0 and CDF Collaborations at the Tevatron measured the ratio of jet-gap-jet events with respect to dijets as a function of jet $p_T$ and $\Delta \eta$ for a fixed gap region between (-1) and (1) in rapidity~\cite{d0jgj}. D0 data are shown in Fig.~\ref{fig2} and are in good agreement with the BFKL NLL calculations as implemented in HERWIG. 
 
 The CMS Collaboration measured recently the same ratio at the LHC energy of 13 TeV for the same fixed gap region~\cite{totemcms} and the results are shown in Fig.~\ref{fig3}. The BFKL predictions are computed using three definitions for the gap implemented in PYTHIA, namely the theoretical one (pure BFKL calculation), the experimental one (no charged particle above 200 MeV in the gap region $-1 < \eta < 1$ as defined by the CMS Collaboration) and the strict gap one (no particle above 1 MeV in the gap region)~\cite{ourpap} for different values of the survival probabilities $S$ indicated in Fig.. There is a clear discrepancy between the CMS measurement and the expectations using the experimental gap definition whereas the strict gap one leads to a good description of data.
 
\begin{figure}
\centerline{%
\includegraphics[width=14.cm]{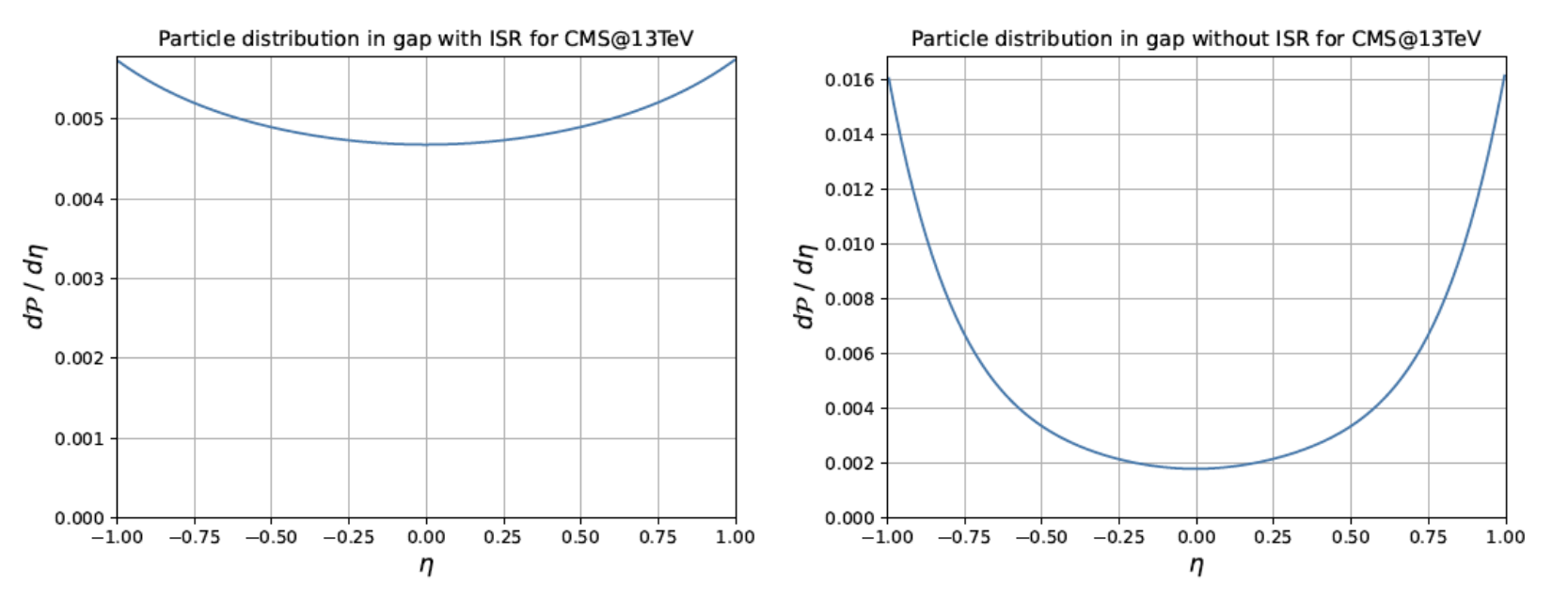}}
\caption{Charged particle production as predicted by PYTHIA in the gap region ($-1<\eta<1$) with ISR on or off.}
\label{fig4}
\end{figure}

 It is thus interesting to understand what changes between the 2 TeV Tevatron and the 13 TeV LHC. It is also worth mentioining that the BFKL calculation also describes the measurements at the LHC at 7 TeV~\cite{ourpap,totemcmsb}. 
 The distribution of charged particles with $P_T>200$ MeV (as defined by the CMS gap definition) from PYTHIA in the gap region $-1<\eta <1$ with initial state radiation  (ISR) ON and OFF are shown respectively on the left and right plots of Fig.~\ref{fig4}~\cite{ourpap}.
Particles emitted at large angle with $p_T > 200$ MeV from ISR have a large influence on the gap presence or not, and thus on the gap definition (experimental or strict). It means that the number of particles emitted in the gap region and predicted by PYTHIA is too large and would need further tuning using data.

The second point to be understood is why the discrepancy between the BFKL calculation and data was mainly at 13 TeV and not present at lower center-of-mass energies. As we mentioned, the ratio between jet gap jets and inclusive jets is measured. The events predicted by the BFKL dynamics using the experimental and strict gap definitions are more quark gluon induced processes at Tevatron energies and  gluon gluon ones at LHC energies. It is the same for inclusive jet production (except  at large $\Delta \eta$ where quark gluon processes dominate at the LHC). 
The number of emitted particles by QCD radiation is much larger for gluon gluon processes than for quark gluon processes, and obviously with ISR ON. The fact that the agreement between BFKL calculations as implemented in PYTHIA and CMS measurements is poor is thus due to two reasons, namely too much ISR in PYTHIA and the fact that gluon gluon processes dominate at the 13 TeV LHC~\cite{ourpap}.
The ISR emission from PYTHIA is too large at high angle and must be further tuned for jet gap jet events using for instance $J/\Psi$-gap-$J/\Psi$ events which is a gluon gluon dominated process.

The full NLO BFKL calculation of jet gap jet processes was recently performed in Ref.~\cite{jgjnlo} including the NLO impact factors. The effects of NLO corrections were found to be quite small and do not change the conclusions concerning ISR radiation in PYTHIA.

\section{First observation of jet gap jet events in diffraction by the CMS Collaboration}

\begin{figure}
\centerline{%
\includegraphics[width=14.cm]{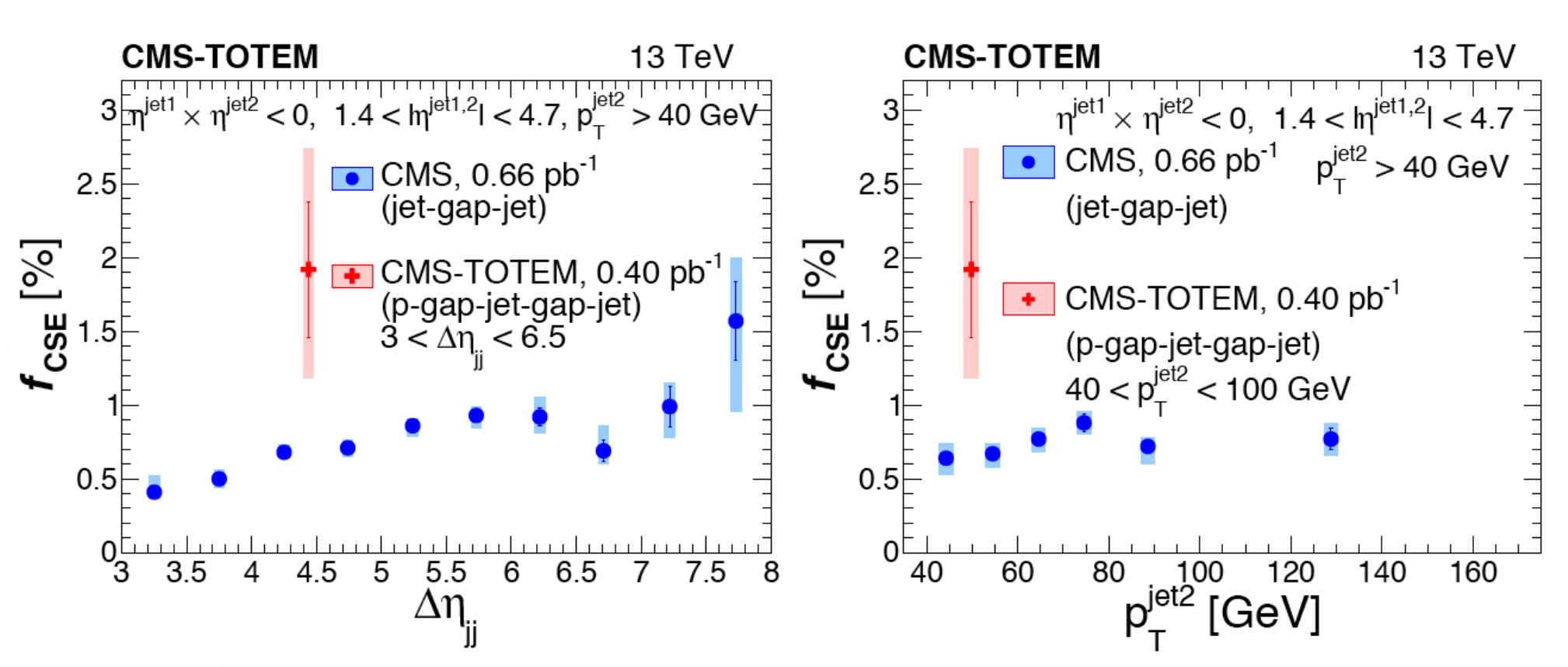}}
\caption{First observation of jet gap jet events in diffraction by the CMS Collaboration.}
\label{fig5}
\end{figure}

The first measurement of jet gap jet events in diffraction was also performed recently by the CMS and TOTEM collaborations~\cite{totemcms}.  These events are very clean since multi-parton interaction effects are suppressed by requesting at least one proton to be tagged~\cite{jgjpap} and could represent an ideal process to look for BFKL resummation. 11 events were observed with a gap between jets and at least one proton tagged with about 
0.7  pb$^{-1}$, as shown in Fig.~\ref{fig5} where we display the jet gap jet fraction as a function of $\Delta \eta$ between the jets and $p_T$ of the second leading jets for diffractive and inclusive events. It is clear that the fraction of jet gap jet events is enhanced in diffraction. This measurement would benefit from more luminosity to get more differential measurements.

To conclude, we presented a measurement of the jet gap jet fraction  at the Tevatron (1.96 TeV) and at the LHC (7 and 13 TeV). A good agreement between the BFKL calculation and the measurement is found at  Tevatron energies, but an apparent disagreement appears at 13 TeV. 
BFKL predictions are in fact very sensitive to ISR as described in PYTHIA especially for gluon gluon interaction processes, that dominate at 13 TeV.
Too much ISR at high angle is predicted by PYTHIA and further  tuning  using for instance $J/\Psi$-gap-$J/\Psi$ events
should be performed.

\end{document}